\documentclass[prl,twocolumn,preprintnumbers,amsmath,amssymb]{revtex4}

\usepackage{graphicx}
\usepackage{dcolumn}
\usepackage{bm}
\usepackage{times}

\begin{document}

\bibliographystyle{unsrt}
\preprint{A. H\"{o}gele {\em et al.}, August 2004}

\title{Voltage-controlled electron-hole interaction in a single quantum dot}

\author{Alexander H\"{o}gele}
\affiliation{Center for NanoScience and Department f\"{u}r Physik, Ludwig-Maximilians-Universit\"{a}t,
Geschwister-Scholl-Platz 1, 80539 M\"{u}nchen, Germany}

\author{Stefan Seidl}
\affiliation{Center for NanoScience and Department f\"{u}r Physik, Ludwig-Maximilians-Universit\"{a}t,
Geschwister-Scholl-Platz 1, 80539 M\"{u}nchen, Germany}

\author{Martin Kroner}
\affiliation{Center for NanoScience and Department f\"{u}r Physik, Ludwig-Maximilians-Universit\"{a}t,
Geschwister-Scholl-Platz 1, 80539 M\"{u}nchen, Germany}

\author{Khaled Karrai}
\affiliation{Center for NanoScience and Department f\"{u}r Physik, Ludwig-Maximilians-Universit\"{a}t,
Geschwister-Scholl-Platz 1, 80539 M\"{u}nchen, Germany}

\author{Mete Atat\"{u}re}
\affiliation{Institute of Quantum Electronics, ETH H\"{o}nggerberg HPT G12, CH-8093, Z\"{u}rich, Switzerland}

\author{Jan Dreiser}
\affiliation{Institute of Quantum Electronics, ETH H\"{o}nggerberg HPT G12, CH-8093, Z\"{u}rich, Switzerland}

\author{Atac Imamo\u{g}lu}
\affiliation{Institute of Quantum Electronics, ETH H\"{o}nggerberg HPT G12, CH-8093, Z\"{u}rich, Switzerland}

\author{Richard J.\ Warburton}
\affiliation{School of Engineering and Physical Sciences, Heriot-Watt University, Edinburgh EH14 4AS, UK}

\author{Brian D. Gerardot}
\affiliation{Materials Department, University of California, Santa Barbara, California 93106, USA}

\author{Pierre M. Petroff}
\affiliation{Materials Department, University of California, Santa Barbara, California 93106, USA}

\date{\today}



\begin{abstract}
The ground state of neutral and negatively charged excitons confined to a single self-assembled InGaAs quantum
dot is probed in a direct absorption experiment by high resolution laser spectroscopy. We show how the
anisotropic electron-hole exchange interaction depends on the exciton charge and demonstrate how the interaction
can be switched on and off with a small dc voltage. Furthermore, we report polarization sensitive analysis of
the excitonic interband transition in a single quantum dot as a function of charge with and without magnetic
field.

\end{abstract}

\maketitle

Spin control and manipulation in mesoscopic semiconductor systems have attracted extensive attention within the
last few years. The activity in this field is driven by the idea of using spin states for quantum information
processing and quantum communication. In particular, semiconductor quantum dots (QDs) have been considered for
realization of spin quantum bits \cite{Loss,Burkard} as they offer the potential advantage of scalability and
tunability. For spin qubit processing in QDs, an optical scheme has been envisioned \cite{Imamoglu}. Other
proposals involve a combination of spin and charge excitation \cite{Chen} or an all-optical implementation of
quantum information processing \cite{Biolatti} in QDs. All proposals have a common crucial requirement, namely
resonant and spin selective excitation.

Significant progress has been made with naturally formed QDs \cite{GammonScience} based on resonant control of
excitonic states \cite{Guest,Stievater}, leading to the recent demonstration of an optical CROT gate \cite{Li}.
Self-assembled QDs have the advantage of longer excitonic coherence time due to stronger confinement. They have
been proved to serve as a source of non-classical light for secure quantum communication
\cite{Michler,Moreau,Yuan,Santori}. An implementation of self-assembled QDs as a spin sensitive post processing
read-out tool can be envisioned. Electric dipole transitions are spin sensitive, such that the spin information
of the optically active state is imprinted onto the photon polarization. High efficiency single photon devices
\cite{Pelton} could provide high yield for spin qubit detection.

Recently, we have reported resonant exciton creation into the ground state \cite{PhysE,PRL} and the first
excited state \cite{Alen} of a single self-assembled QD. In the work presented here, we address the topic of
polarization selective resonant creation of excitonic states in a single self-assembled InGaAs QD by high
resolution laser spectroscopy. We report results on the spin mediated anisotropic electron-hole exchange and on
the polarization dependence of the excitonic states as function of charge, electric and magnetic field.

The InGaAs QDs investigated in the experiments were grown by molecular beam epitaxy in the self-assembly
Stranski-Krastanow mode and are embedded in a field effect heterostructure \cite{Drexler}. Highly n-doped GaAs
acts as back electrode followed by a tunnel barrier of 25nm GaAs and the InGaAs QDs. An annealing step was
introduced in order to shift the photoluminescence (PL) emission energy to around 1.3~eV \cite{Garcia}. The QDs
are sequentially capped with 30~nm GaAs and a 120~nm AlAs/GaAs superlattice. A semitransparent NiCr gate
electrode evaporated on the surface allows us to control the excitonic properties of QDs by applying a voltage
with respect to the back contact. The exciton energy can be fine tuned using gate voltage induced vertical Stark
effect \cite{Alen}. Furthermore, the QDs can be charged sequentially with electrons from the metallic-like back
electron reservoir. For a single QD the charge state is unambiguously identified by monitoring pronounced
Coulomb blockade in the PL \cite{Nature,Khaled}.

\begin{figure}[t]
\begin{center}
\includegraphics[scale=0.95]{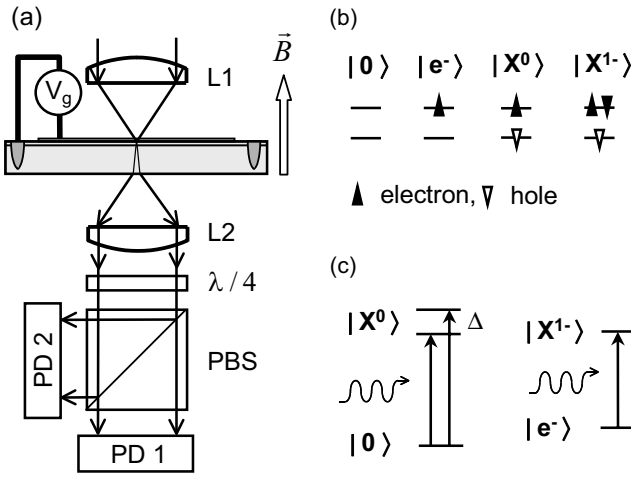}
\end{center}
\caption{\label{fig1}(a) Optical transmission setup: Tunable narrow band laser light is delivered with an
optical fiber (not shown), collimated and then focussed with the aspherical lens L1 with numerical aperture 0.55
onto the sample. The transmitted light is collimated with the lens L2. Before detection with the Ge p-i-n
photodiodes (PD 1, PD 2), the transmitted light passes a quarter-wave plate ($\lambda /4$) and a polarizing
beamsplitter (PBS). The charge state of quantum dots is defined by gate voltage V$_{g}$, the magnetic field $B$
is applied in Faraday configuration perpendicular to the sample surface. (b) Quantum mechanical states in a
single quantum dot: $| 0 \rangle$ is the vacuum state, $| {\rm e}^{-}\rangle$ the single electron state, $| {\rm
X}^{0}\rangle$ the neutral exciton state, and $| {\rm X}^{1-}\rangle$ the singly charged exciton state. (c) The
level diagrams for the optical creation of a neutral exciton and a singly charged exciton. The neutral exciton
is split by $\Delta$ through the anisotropic electron-hole exchange interaction.}
\end{figure}

We used a home built fiber-based confocal microscope for both the PL and the differential transmission
spectroscopy (Fig.~1a). For all experiments presented here, the microscope was cooled to liquid Helium
temperature in a bath cryostat. Details of the experimental setup have been discussed elsewhere \cite{PhysE}.
Excitation laser light was provided through a single-mode glass fiber, collimated and focused on the sample
surface with a lens L1 with numerical aperture of 0.55. The sample was brought into the focal plane with a low
temperature compatible XYZ-positioning stage (Atto Cube Systems, ANP-XYZ-100), allowing for precise vertical and
lateral positioning. For most experiments, a commercial Ge photodiode was sandwiched between the sample and the
positioning stage in order to detect the total laser light transmitted through the sample. Two types of Ge
infrared photodetectors were used: FDG05, Thorlabs Inc. and J16-SC, EG\&G Judson (J16-C11-R02M-SC) with an
active diameter of 5~mm and 2~mm, respectively. The advantage of the latter photodiode is a factor of 5 lower
noise and a factor of 3 higher bandwidth at 4.2~K. Alternatively, the photodetector was replaced by a
polarization analysis setup. It is well known that fiber bending can modify the light polarization. However, by
adjusting the degree of polarization with a combination of half- and quarter-wave plates before coupling the
laser light into the microscope fiber, any fiber polarization contribution can be compensated. Alternatively,
fiber bending paddles could be used for polarization control. The light reflected from the sample surface and
detected outside the cryostat would then provide information on the degree of polarization. In our experiments
however, we used the polarization analysis setup as shown in Fig.~\ref{fig1}a in order to identify unambiguously
the polarization state of the light focused onto the sample.

\begin{figure}[t]
\begin{center}
\includegraphics[scale=0.95]{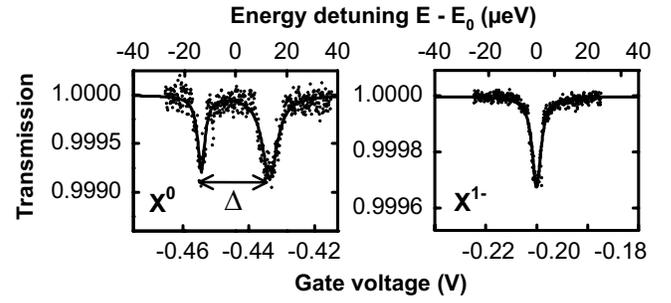}
\end{center}
\caption{\label{fig2}Differential transmission of the neutral X$^{0}$ (left) and charged exciton X$^{1-}$
(right) in a single self-assembled quantum dot. The detuning was achieved at constant laser wavelength by
sweeping the gate voltage. The two resonances of the neutral exciton are split by the fine structure
$\Delta=27~\mu$eV. The resonance energy E$_{0}$ of X$^{0}$ was 1.272~eV and that of X$^{1-}$ was 1.266~eV. The
sample was at 4.2~K, no magnetic field was applied.}
\end{figure}

The polarization analysis setup contains an additional lens L2 (Geltech Aspheric Lens, 350230-B) which
collimates the laser light. The parallel beam passes a quarter-wave plate (CVI, QWPO-950-04-4) with the fast
axis oriented with an angle of $-$45$^{\circ}$ with respect to the $p$-axis of the polarizing cube beamsplitter
PBS (CVI, PBS-930-020). The magnetic field is aligned parallel to the sample growth direction and antiparallel
to the propagation $k$-vector of the excitation beam (Fig.~\ref{fig1}). Given this orientation, the quarter-wave
plate transforms $\sigma ^{-}$ and $\sigma ^{+}$ circular light into linearly $s$-polarized and $p$-polarized
light, respectively. The $p$-polarized component is transmitted onto a Ge p-i-n photodiode PD~1
(J16-C11-R02M-SC, EG\&G Judson) whereas the $s$-component is directed to the photodiode PD~2 of the same type.
The signal intensity detected by the photodiodes is anti-correlated and sums up to the total signal of the
transmitted light. In order to avoid losses, the parameters of the lens L2 were chosen such that the waist of
the collimated beam does not exceed the active area of the photodetectors. The ratio of the signals on PD~1 and
PD~2 allows for a direct determination of the degree of ellipticity of the excitation light. In particular, in
the case of pure right-hand circular polarization, the detector PD 1 shows maximum signal whereas the signal on
PD 2 is minimal. For left-hand circular polarization, the situation is reversed with minimal signal on PD 1 and
maximal signal on PD 2. In order to analyze linear polarization, the quarter-wave plate had to be removed. Prior
to application in the spectroscopy experiments, we tested the polarization analysis setup and we confirmed that
it operated at room temperature as well as at liquid Nitrogen and liquid Helium temperatures.

Fig.~\ref{fig1}b shows schematically the quantum mechanical states in a single QD probed by means of both
resonant and non-resonant spectroscopy. With non-resonant PL spectroscopy, we first identify the exciton
energies and the gate voltage regions of the neutral exciton X$^{0}$ and the charged exciton X$^{1-}$ in a
single QD \cite{Nature}. As observed in PL on several dozens QDs emitting around 1.3~eV, the typical
corresponding voltage intervals for low excitation power are $\lbrack -0.8$~V,~$-0.4~$V\,$\rbrack$ and $\lbrack
-0.4$~V,~$-0.1~$V\,$\rbrack$ for the X$^{0}$ exciton and the X$^{1-}$ exciton, respectively. Then, a narrow band
tunable diode laser (Sacher Lasertechnik, TEC500, $\Delta \nu \leq$5~MHz) is adjusted to match the energy of the
interband optical transition into the neutral or singly charged excitonic ground state of the selected QD
(Fig.~\ref{fig1}c). The detuning of the transition energy with respect to the laser excitation energy is
achieved through the Stark effect by sweeping the gate voltage \cite{Alen}. The Stark shift depends
quadratically on the applied voltage \cite{WarburtonStark} but in a small range of gate voltage it can be
approximated by a linear function. In the present case, the typical Stark shift is $\sim $1~meV/V and does not
depend on the charge state of the exciton.

\begin{figure}
\begin{center}
\includegraphics[scale=1]{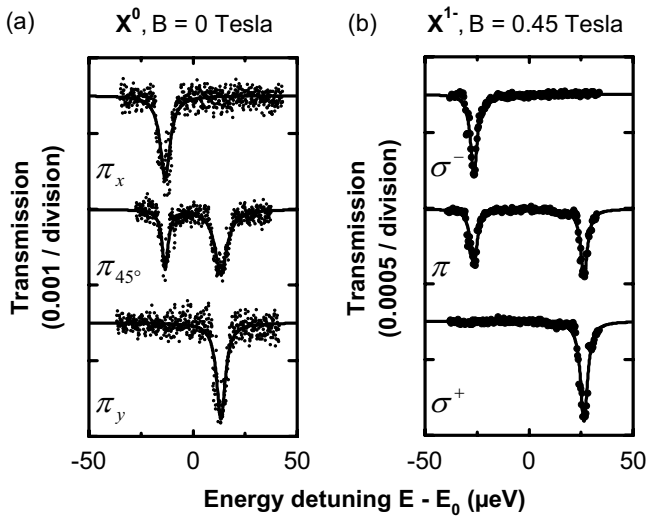}
\end{center}
\caption{\label{fig3}Polarization dependence of the optical transitions in a single quantum dot$^{*}$ at 4.2~K.
(a) Neutral exciton X$^{0}$ spectra for three different linear polarizations $\pi_{x}$, $\pi_{45^{\circ}}$ and
$\pi_{y}$ at zero magnetic field. (b) Singly charged exciton X$^{1-}$ spectra for left-hand circular
polarization $\sigma^{-}$, linear polarization $\pi$ and right-hand circular polarization $\sigma^{+}$. The
magnetic field was 0.45 Tesla. In (a) and (b) the curves were offset vertically for clarity. $^{*}$Note:
Experimental data in Fig.s~2, 3 a, in Fig.s~3 b, 4 right and in Fig. 4 left were recorded on three different
quantum dots.}
\end{figure}

Differential transmission spectra within the corresponding gate voltage interval of the neutral and charged
exciton in a single QD at zero magnetic field are shown in Fig.~\ref{fig2}. The X$^{0}$ transition, resonant
with excitation laser energy $E_{0}=1.272$~eV, exhibits two lines split by the fine structure $\Delta =$
27~$\mu$eV. In contrast, the X$^{1-}$ exhibits only a single resonance. The linewidths of the resonances in
Fig.~\ref{fig2} are 3.5~$\mu$eV and 7.1~$\mu$eV for the $\pi_{x}$ and $\pi_{y}$ transition of the neutral
exciton and 4.2~$\mu$eV for the X$^{1-}$ transition. It is not always the case that the $\pi_{y}$ resonance is
broader than the $\pi_{x}$ resonance. From time resolved measurements on single QDs in a similar sample we
expect the linewidth to be $\sim$1~$\mu$eV for neutral and charged excitons. However, we find that the exciton
energy experiences spectral fluctuation of several $\mu$eV which broadens the resonance line \cite{PRL}.

The interband transition energy of the charged exciton in Fig.~2 is 6~meV below the X$^{0}$ energy, a
consequence of the difference in binding energy \cite{Nature}. The fine structure arises through electron-hole
exchange interaction in a QD potential with reduced symmetry \cite{GammonFS}. A splitting of several $\mu$eV is
expected even for cylindrically symmetric QDs due to the lack of inversion symmetry in the underlying lattice
\cite{Zunger}. For In$_{x}$Ga$_{1-x}$As QDs, the splitting can be as high as 200~$\mu$eV for strongly asymmetric
dots \cite{BayerFS}; Langbein et al. report a decrease of the fine structure splitting down to 6~$\mu$eV with
annealing \cite{Langbein}. In our sample, the value of the splitting varies from 11~$\mu$eV to 42~$\mu$eV as
measured on several individual QDs. The magnitude of $\Delta$ indicates that the dominant contribution arises
through QD shape anisotropy. We are able to switch off the spin mediated electron-hole exchange by applying a
small dc voltage. In the charged exciton state, the two electrons have opposite spins (Fig.~\ref{fig1}b) and the
total electron spin is zero. For this reason, the electron-hole exchange interaction vanishes and no splitting
is observed (Fig.~\ref{fig2}~left).

\begin{figure}
\begin{center}
\includegraphics[scale=1]{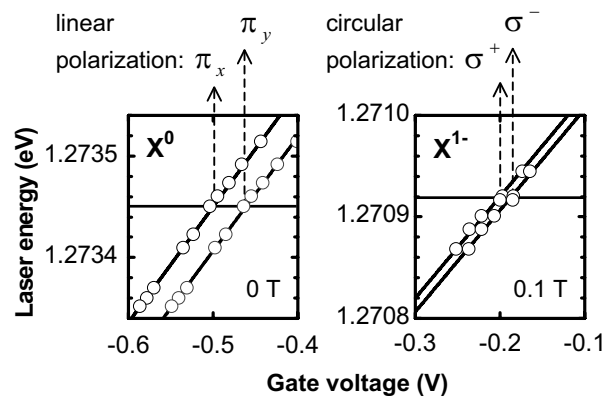}
\end{center}
\caption{\label{fig4} Voltage control of polarization in a single quantum dot. Resonance energies of linearly
polarized X$^{0}$ transitions at zero magnetic field (left) and circularly polarized X$^{1-}$ transitions at 0.1
Tesla (right) as function of gate voltage. The Zeeman splitting is 12~$\mu$eV. The data were taken at 4.2~K.}
\end{figure}

Fig.~3 shows the polarization characteristics of the neutral and charged exciton resonances. At zero magnetic
field, the two X$^{0}$ states are expected to couple to photons having orthogonal linear polarizations. The
experimental results shown in Fig.~\ref{fig3}a confirm the picture: the absorption resonances are sensitive only
to photons with appropriate linear polarization. A magnetic field applied in the growth direction splits the
X$^{1-}$ transition into two lines separated by the Zeeman energy $E_{Z}=g^{*}\mu _{B}B$, where $g^{*}$ is the
exciton $g$-factor and $\mu _{B}$ the Bohr magneton. In our sample, a typical value of $g^{*}$ is $-$2 leading
to a characteristic Zeeman splitting of $\sim$ 120~ $\mu$eV/\,T$^{2}$ \cite{Schulhauser}. Fig.~\ref{fig3}b shows
the polarization dependence of the two Zeeman split X$^{1-}$ lines in a magnetic field of 0.45 Tesla. For linear
polarization $\pi$ both resonances are active. Each transition can be addressed individually with circularly
polarized light. The low and high energy branch were found to be sensitive to the orthogonal circular
polarizations $\sigma ^{-}$ and $\sigma ^{+}$, respectively. This is anticipated for a negative exciton
$g$-factor in agreement with earlier reports for similar samples \cite{BayerFS, BayerEH}. For magnetic field
higher than 6 Tesla the $\sigma ^{-}$ resonance was strongly inhibited and $\sigma ^{+}$ resonance became
dominant \cite{APL}.

As a consequence of the results described above, the optical polarization property of an individual QD can be
controlled by the dc voltage applied between the top and the back electrode. For a given photon energy at zero
magnetic field the QD's absorption within the voltage interval of the neutral exciton can be switched between
the two orthogonal linear polarizations (Fig.~\ref{fig4}~left). A small dc voltage is sufficient in order to
switch in between the base vectors of the linear polarization. Furthermore, by applying a small magnetic field
and thereby splitting the charged exciton into spin-polarized Zeeman branches, orthogonal circular transitions
are optically active in a single QD (Fig.~\ref{fig4}~right). Again, a small voltage change is necessary in order
to address the two orthogonal circular polarization states but to keep the resonance energy fixed. One should
keep in mind, however, that charging the dot with a single electron shifts the resonance energy by 6~meV. The
voltage-controlled polarization selection scheme is also valid for the photon emission, a very attractive
application for QDs as a source of single photons with switchable polarization bases.

In summary, we have demonstrated that a self-assembled QD can be used to prepare excitonic states with an
unprecedented degree of tunability. The tuning is achieved simply through a voltage. This property was
demonstrated by applying high resolution laser spectroscopy to a single QD. Our results demonstrate that the
electron-hole spin exchange interaction can be switched off in a controlled way with gate voltage by adding a
resident electron to a single QD through field effect. As a consequence of our results, the polarization of
optical emission from a single QD can be switched between linear and circular polarization bases with dc voltage
and a small applied magnetic field, an attractive feature for QDs in single photon source applications.

Financial support for this work was provided in Germany by DFG grant no.\ SFB~631, in Switzerland by NCCR
Quantum Photonics grant no. 6993.1 and in the UK by the EPSRC.


\begin{thebibliography}{}

\bibitem{Loss} D. Loss and D. P. DiVincenzo, Phys. Rev. A {\bf 57}, 120 (1998).

\bibitem{Burkard} G. Burkard, D. Loss, and D. P. DiVincenzo, Phys. Rev. B {\bf 59}, 2070 (1999).

\bibitem{Imamoglu} A. Imamo\u{g}lu, D. D. Awschalom, G. Burkard, D. P. DiVincenzo, D. Loss, M. Sherwin, and A. Small, Phys. Rev. Lett. {\bf 83}, 4240 (1999).

\bibitem{Chen}  P. Chen, C Piermarocchi, and L. J. Sham, Phys. Rev. Lett. {\bf 87}, 067401 (2001).

\bibitem{Biolatti} E. Biolatti, R. C. Iotti, P. Zanardi, and F. Rossi, Phys. Rev. Lett. {\bf 85}, 5647 (2000).

\bibitem{GammonScience} D. Gammon, E. S. Snow, B. V. Shanabrook, D. S. Katzer, and D.
Park, Phys. Rev. Lett. {\bf 76}, 3005 (1996).

\bibitem{Guest} J. R. Guest {\em et al.}, Phys. Rev. B {\bf 65}, 241310 (2002).

\bibitem{Stievater} T. H. Stievater, X. Li, J. R. Guest, D. G. Steel, D. Gammon, D. S. Katzer, and D. Park, Appl. Phys. Lett. {\bf 80}, 1876 (2002).

\bibitem{Li} X. Li, Y. Wu, D. Steel, D. Gammon, T. H. Stievater, D. S. Katzer,
D. Park, C. Piermarocchi, and L. J. Sham, Science {\bf 301}, 809 (2003).

\bibitem{Michler} P. Michler, A. Kiraz, C. Becher, W. V. Schoenfeld, P. M. Petroff,
L. Zhang, E. Hu, and A. Imamo\u{g}lu, Science {\bf 290}, 2282 (2000).

\bibitem{Moreau} E. Moreau, I. Robert, L. Manin, V. Thierry-Mieg, J. M. G\'{e}rard,
I. Abram, Phys. Rev. Lett. {\bf 87}, 1836011 (2001).

\bibitem{Yuan} Z. Yuan, B. E. Kardynal, R. M. Stevenson, A. J. Shields, C. J.
Lobo, K. Cooper, N. S. Beattie, D. A. Ritchie, and M. Pepper, Science {\bf 295}, 102 (2001).

\bibitem{Santori} C. Santori, D. Fattal, J. Vu\u{c}kovi\'{c}, G. S. Solomon, and Y.
Yamamoto, Nature {\bf 419}, 594 (2002).

\bibitem{Pelton} M. Pelton, C. Santori, J. Vu\u{c}kovi\'{c}, B. Zhang, G. S. Solomon, J. Plant, and Y. Yamamoto, Phys. Rev. Lett. {\bf 89}, 233602 (2002).


\bibitem{PhysE} A. H\"{o}gele, B. Al\`{e}n, F. Bickel, R. J. Warburton, P. M.
Petroff, and K. Karrai, Physica E {\bf 21}, 175 (2004).

\bibitem{PRL} A. H\"{o}gele, S. Seidl, M. Kroner, R. J. Warburton, K. Karrai, B. D. Gerardot, and P. M.
Petroff, cond-mat/0408089 (2004).

\bibitem{Alen}
B. Al\`{e}n, F. Bickel, K. Karrai, R. J. Warburton, and P. M. Petroff, Appl. Phys. Lett. {\bf 83}, 2235 (2003).

\bibitem{Drexler} H. Drexler, D. Leonard, W. Hansen, J. P. Kotthaus, and P. M. Petroff, Phys. Rev. Lett. {\bf 73}, 2252 (1994).

\bibitem{Garcia} J. M. Garcia, G. Medeiros-Ribeiro, K. Schmidt, T. Ngo, J. L. Feng, A. Lorke, J. P. Kotthaus, and P. M. Petroff, Appl. Phys. Lett. {\bf 71}, 2014 (1997).

\bibitem{Nature}
R. J. Warburton, C. Sch\"{a}flein, D. Haft, A. Lorke, K. Karrai, J. M. Garcia, W. Schoenfeld, and P. M. Petroff,
Nature {\bf 405}, 926 (2000).

\bibitem{Khaled}
K. Karrai, R. J. Warburton, C. Schulhauser, A. H\"{o}gele, B. Urbaszeck, E. J. McGhee, A. O. Govorov, J. M.
Garcia, B. D. Gerardot, and P. M. Petroff, Nature {\bf 427}, 135 (2004).

\bibitem{WarburtonStark} R. J. Warburton, C. Schulhauser, D. Haft, C. Sch\"{a}flein, K. Karrai, J. M. Garcia, W. Schoenfeld, and P. M. Petroff, Phys. Rev. B {\bf 65}, 113303 (2002).

\bibitem{GammonFS} D. Gammon, E. S. Snow, B. V. Shanabrook, D. S. Katzer, and D. Park, Phys. Rev. Lett {\bf 76}, 3005 (1996).

\bibitem{Zunger} G. Bester, S. Nair, and A. Zunger, Phys. Rev. B {\bf 67}, 161306 (2003).

\bibitem{BayerFS} M. Bayer, A. Kuther, A. Forchel, A. Gorbunov, V. B. Timofeev, F. Sch\"{a}fer, J. P. Reithmaier, T. L. Reinecke, and S. N. Walck, Phys. Rev. Lett. {\bf 82}, 1748 (1999).

\bibitem{Langbein} W. Langbein, P. Borri, U. Woggon, V. Stavarache, D. Reuter, and A. D. Wieck, Phys. Rev. B {\bf 69}, 161301 (2004).

\bibitem{Schulhauser} C. Schulhauser, D. Haft, R. J. Warburton, K. Karrai, A. O. Govorov, A. V. Kalameitsev, A. Chaplik, W. Schoenfeld, J. M. Garcia, and P. M. Petroff, Phys. Rev. B {\bf 66}, 193303 (2002).

\bibitem{BayerEH} M. Bayer {\em et al.}, Phys. Rev. B {\bf 65}, 195315 (2002).

\bibitem{APL} A. H\"{o}gele, M. Kroner, S. Seidl, K. Karrai, M. Atat\"{u}re, J. Dreiser, A. Imamo\u{g}lu, R. J. Warburton, B. D. Gerardot, and P. M. Petroff, to be published (2004).

\end{thebibliography}
\end{document}